\documentclass[journal]{IEEEtran}

\usepackage{graphicx}
\usepackage{amssymb}
\usepackage{multirow}
\usepackage{caption}
\usepackage{subcaption}
\newcommand\norm[1]{\left\lVert#1\right\rVert}

\usepackage{tikz}
\usetikzlibrary{shapes,arrows,positioning,calc,fit}
\usepackage{amsmath}

\DeclareMathOperator*{\argmax}{arg\,max}
\DeclareMathOperator*{\argmin}{arg\,min}

\def\BibTeX{{\rm B\kern-.05em{\sc i\kern-.025em b}\kern-.08em
    T\kern-.1667em\lower.7ex\hbox{E}\kern-.125emX}}
\ifCLASSINFOpdf
\else
\fi
\begin{document}
%
\title{Deep Proximal Learning for High-Resolution Plane Wave Compounding}
%
%
%

\author{Nishith~Chennakeshava,~\IEEEmembership{Student~Member,~IEEE,}
        Ben~Luijten,~\IEEEmembership{Student~Member,~IEEE,}
        Massimo~Mischi,~\IEEEmembership{Senior~Member,~IEEE,}
        Yonina~C.~Eldar,~\IEEEmembership{Fellow,~IEEE}
        and~Ruud~J.~G.~van~Sloun,~\IEEEmembership{Member,~IEEE}
\thanks{}
\thanks{}
\thanks{}}

\maketitle

\begin{abstract}


Plane Wave imaging enables many applications that require high frame rates, including localisation microscopy, shear wave elastography, and ultra-sensitive Doppler. To alleviate the degradation of image quality with respect to conventional focused acquisition, typically, multiple acquisitions from distinctly steered plane waves are coherently (i.e. after time-of-flight correction) compounded into a single image. This poses a trade-off between image quality and achievable frame-rate. To that end, we propose a new deep learning approach, derived by formulating plane wave compounding as a linear inverse problem, that attains high resolution, high-contrast images from just 3 plane wave transmissions. Our solution unfolds the iterations of a proximal gradient descent algorithm as a deep network, thereby directly exploiting the physics-based generative acquisition model into the neural network design. We train our network in a greedy manner, i.e. layer-by-layer, using a combination of pixel, temporal, and distribution (adversarial) losses to achieve both perceptual fidelity and data consistency. Through the strong model-based inductive bias, the proposed architecture outperforms several standard benchmark architectures in terms of image quality, with a low computational and memory footprint. 

\end{abstract}


%
\IEEEpeerreviewmaketitle

\section{Introduction}\label{chapter:introduction}

\begin{figure*}[t!]
    \centering
    \includegraphics[trim={0 11cm 4 3cm},clip,width=\textwidth]{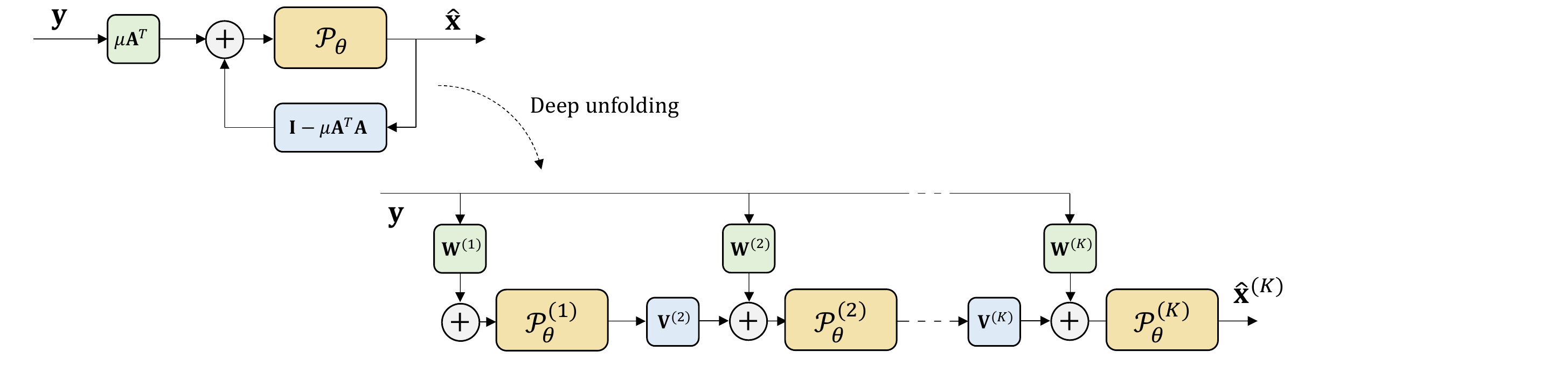}
    \caption{Unfolded deep proximal learning, obtained by unfolding the recurrent proximal gradient scheme to a K layered neural network.}
    \label{fig:network}
\end{figure*}

In Plane Wave (PW) imaging, multiple low resolution images are acquired using steered transmissions, which are then coherently compounded to obtain a high quality image with enhanced lateral resolution and contrast. Compounding more steered plane wave acquisitions improves image quality \cite{tanter2014ultrafast}, but comes at the cost of temporal resolution, i.e. frame-rate. This poses challenges for applications that rely on high frame rates, such as elastography, Ultrasound Localisation Microscopy (ULM) \cite{van2020super}, or sensitive Doppler. One way to improve temporal resolution is to reduce the number of steered PWs transmitted per acquisition, but this negatively impacts lateral resolution. In this paper, we propose a model-based Deep Learning (DL) solution that aims to relax this trade-off between lateral resolution and frame-rate, achieving high resolution and contrast by compounding just 3 PW transmissions.

Deep Learning has revolutionised many domains. It is also increasingly being used in ultrasound based applications \cite{van2019deep, bell2020challenge}. Convolutional Neural Nets (CNNs), such as ResNets \cite{szegedy2017inception} and UNets \cite{ronneberger2015u} have produced excellent results in denoising tasks \cite{wang20183d}. Other applications of Deep Neural Networks (DNNs) in ultrasound include the weighting of channel signals (adaptive beamforming by deep learning) \cite{luijten2020adaptive}, probabilistic sub-sampling for compressed sensing \cite{huijben2019deep}, and coherent PW compounding \cite{gasse2017high}. Related to this work, Gasse et al. (2017) \cite{gasse2017high} use a CNN in tandem with Maxout activation functions, to coherently compound 3 beamformed images, acquired using 1 PW each, and map them towards an image acquired using 31 PWs. Most of the work in the domain of plane wave compounding is achieved by using a variation of the ResNet \cite{khan2019universal} or the UNet architectures \cite{guo2020high}, sometimes in combination with a discriminator in a Generative Adversarial Network-like (GAN) training strategy \cite{zhang2018high, goodfellow2014generative}. There are also works exploring channel-based compounding, in pursuit of better image quality \cite{rothlubbers2020improving}. However, generic convolutional neural networks like a UNet, or a ResNet may sometimes yield solutions that are either physically invalid, or are visually implausible. In other words, they may violate the physical measurement model. Such Neural Networks (NNs) are also typically over-parameterised \cite{cohen2018distribution}. By training light weight models, we tend to limit or avoid the under-specification problem, as there are fewer degrees of freedom in the NN architecture \cite{d2020underspecification, geirhos2020shortcut, cohen2019deep}.

This study expands upon the initial work of \cite{chennakeshava2020high}, by incorporating physics-based inductive biases. We do this by employing physics-based signal priors in the design of our NN, such that we may obtain a network that is compact, yet performs very well when compared to benchmarks (a UNet, ResNet, and \cite{gasse2017high}). In addition, we take further advantage of the signal's noise characteristics in this work. Therefore, this problem is formulated as an ill-posed, linear inverse problem that we solve using data-driven techniques. We achieve an additional boost in the spatial resolution and contrast of the resulting image by training towards images acquired using a higher frequency, and many more PWs. Furthermore, to promote consistency between frames we implement a frame-to-frame loss, which suppresses uncorrelated noise between the frames.

We begin by describing the methods employed in deriving the unfolded proximal gradient descent scheme in Section \ref{chapter:methods}, followed by details about the data acquisition technique, and the volume of data collected in Section \ref{chapter:data_acquisition}. We then describe the training strategy in Section \ref{chapter:training}, and showcase the results in Section \ref{chapter:results}. We discuss the results in Section \ref{chapter:discussion}, and finally conclude the paper in Section \ref{chapter:conclusion}.

\begin{figure*}[t!]
    \centering
    \includegraphics[trim={0cm 8cm 2cm 3cm},clip,width=\textwidth]{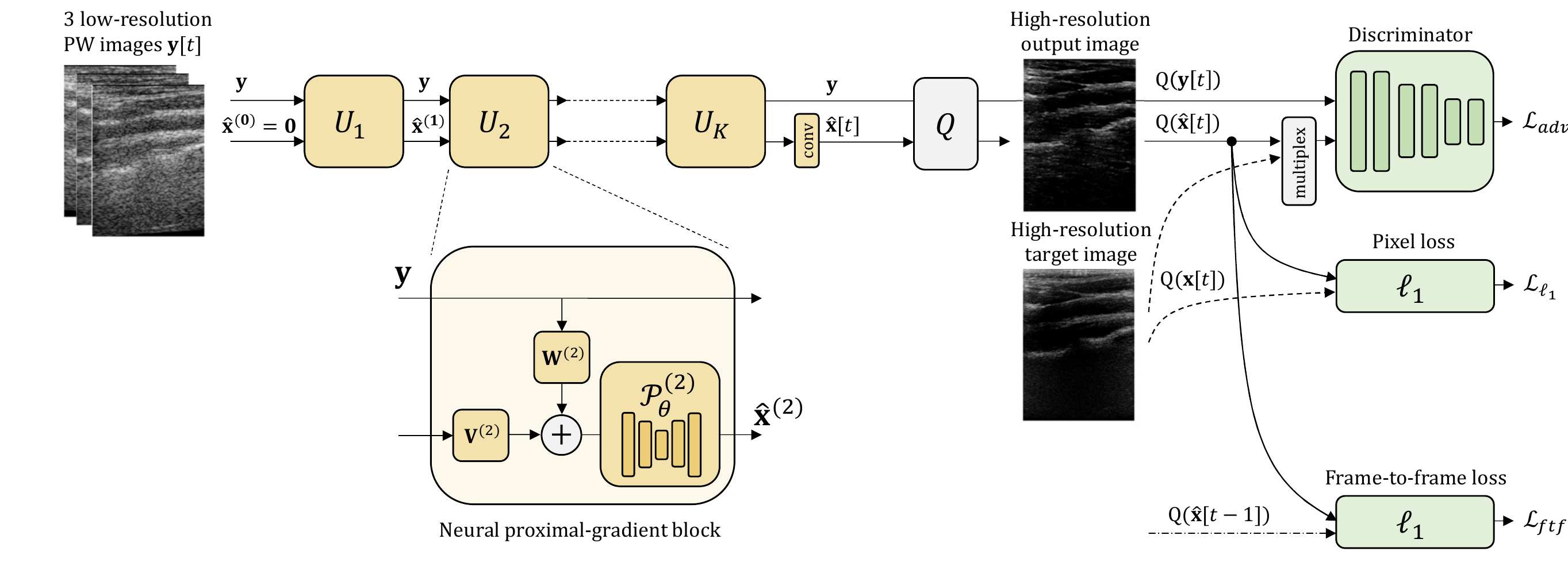}
    \caption{Illustration of data and loss in the training loop. Pink indicates the flow of loss values, and black indicates the flow of data, from input to output. The $U$ blocks indicate the unfoldings of the proximal network, up to $K$ unfoldings. The block to the right hand side of the image with a neuron indicates the discriminator. As we need frame $\textit{t-1}$ and \textit{t} for the calculation of all loss terms, the graphic also displays the training loop across frames in an acquisition.}
    \label{fig:train}
\end{figure*}

\section{Methods}\label{chapter:methods}


\subsection{Signal model and problem formulation}

\noindent We model the problem of compounding multiple plane waves as an inverse problem in which we aim to recover an underlying high-resolution (HR), high-contrast image from a set of low-resolution (LR) noisy measurements with multiple angled plane waves. We denote $\mathbf{x} \in \mathbb{R}^{N}$ as the vectorised high-resolution beamformed RF image, and $\mathbf{y} \in \mathbb{R}^{NM}$ the vectorised measurement of low-resolution beamformed RF images from $M$ transmitted plane waves. We consider these measurements to be acquired according to the following linear model:
\begin{equation}
    \mathbf{y} = \mathbf{A} \mathbf{x} + \mathbf{n}.
    \label{eq:measurement_model}
\end{equation}
Here,
\begin{equation}
    \mathbf{A} = \begin{pmatrix}\mathbf{A}_{1} \\
    \mathbf{A}_{2} \\
    \vdots \\ \mathbf{A}_{M}\end{pmatrix},
    \label{eq:observation_m}
\end{equation}
and
\begin{equation}
    \mathbf{y} = \begin{pmatrix}\mathbf{y}_{1} \\
    \mathbf{y}_{2} \\
    \vdots \\ \mathbf{y}_{M}\end{pmatrix},
    \label{eq:measured_m}
\end{equation}
where $\mathbf{y}_{m}$ is the vectorised, beamformed RF image belonging to the $m^{\textrm{th}}$ steered plane wave transmission, $\mathbf{n}$ $\in$ $\mathbb{R}^{NM}$ is a noise vector which is assumed to follow a Gaussian distribution with zero mean and diagonal covariance, and $\mathbf{A}$ $\in$ $\mathbb{R}^{NM \times N}$ is a block matrix, with its blocks $\mathbf{A}_{1}$, $\mathbf{A}_{2}$,..., $\mathbf{A}_{M}$ being the measurement matrices of individual PW acquisitions. These measurement matrices capture the transformation between our underlying high-resolution RF image $\mathbf{x}$ and each of the plane wave measurements, and are a function of their respective point spread functions. We assume that they follow a convolutional Toeplitz structure through which we may re-write application of $\mathbf{A}_{M}$ as a convolution with an anisotropic `blurring' kernel. 

Solving equation \eqref{eq:measurement_model} for $\mathbf{x}$ is an ill-posed problem, and hence direct maximum-likelihood estimation may lead to solutions that are not consistent with our prior belief of what HR images look like (anatomical and visual plausibility). To incorporate such priors, we estimate $\mathbf{x}$ through Maximum A-Posteriori (MAP) estimation:
\begin{equation}
    \hat{\mathbf{x}} := \argmax_\mathbf{x} p(\mathbf{x}|\mathbf{y}) \propto \argmax_\mathbf{x} p(\mathbf{y}|\mathbf{x}) p_{\theta}(\mathbf{x}),
    \label{eq:map}
\end{equation}
where $\mathbf{\hat{\mathbf{x}}}$ is the estimated high-resolution ultrasound image, and $p(\mathbf{y}|\mathbf{x})$ is the likelihood according to the linear measurement model in \eqref{eq:measurement_model}, having a Gaussian distribution on the model errors. Here, $p_{\theta}(\mathbf{x})$ is a probability density function that expresses prior beliefs about the distribution of HR images $\mathbf{x}$, where $\theta$ are learnt parameters.

Under a Gaussian model on $p(\mathbf{y}|\mathbf{x})$, taking the log of \eqref{eq:map} results in:
\begin{equation}
    \hat{\mathbf{x}} = \argmin_\mathbf{x}\norm{\mathbf{y}-\mathbf{A}\mathbf{x}}_{2}^{2} - \log p_{\theta}(\mathbf{x})
    \label{eq:formal_problem}
\end{equation}
where $\log p_{\theta}(\mathbf{x})$ acts as a regulariser $R(\mathbf{x})$ for the optimisation problem.



\subsection{Proximal gradient iterations}

Assuming $p_{\theta}(\mathbf{x})$ is given, we proceed by formulating a proximal gradient solver for \eqref{eq:formal_problem}. This results in an iterative algorithm that alternates between a gradient update step with respect to the measurement model (data consistency), and a proximal step that pushes the intermediate solutions in the proximity of the regulariser. One iteration is given by:
\begin{equation}
    \begin{aligned}
        \mathbf{\hat{x}}^{(k+1)} &= \mathcal{P}(\mathbf{\hat{x}}^{(k)}- \mu \mathbf{A}^{T}(\mathbf{A}\mathbf{\hat{x}}^{(k)} - \mathbf{y})),
    \end{aligned}
    \label{eq:iterations}
\end{equation}
where $\mathcal{P}^{(k)}$ is the proximal operator of the regularizer $R(\mathbf{x})$ norm \cite{van2019deep}, and $\mu$ is a step size. We can rewrite \eqref{eq:iterations} in a compact form, separating contributions of $\hat{\mathbf{x}}$ and $\mathbf{y}$, as 
\begin{equation}
    \begin{aligned}
        \mathbf{\hat{x}}^{(k+1)} &= \mathcal{P}_{\theta}^{(k)}(\mathbf{\hat{x}}^{(k)}- \mu^{(k)} \mathbf{A}^{T}(\mathbf{A}\mathbf{\hat{x}}^{(k)} - \mathbf{y})) \\
        &= \mathcal{P}_{\theta}^{(k)}(\mu^{(k)} \mathbf{A}^{T}\mathbf{y}+   (\mathbf{I} - \mu^{(k)} \mathbf{A}^{T}\mathbf{A}) \mathbf{\hat{x}}^{(k)}) \\
        &= \mathcal{P}_{\theta}^{(k)}(\mathbf{W}^{(k)}\mathbf{y}+\mathbf{V}^{(k)}\mathbf{\hat{x}}^{(k)}).
    \end{aligned}
\end{equation}
where we denote $\mathbf{W}^{(k)} = \mu^{(k)} \mathbf{A}^{T}$ and $\mathbf{V}^{(k)} = \mathbf{I} - \mu^{(k)} \mathbf{A}^{T}\mathbf{A}$.

For particular distributions $p_{\theta}(\mathbf{x})$ the proximal operator $\mathcal{P}_{\theta}$ has a closed form. For example, if $p_{\theta}(\mathbf{x})$ is a Laplace distribution, equivalent to $\ell_{1}$ regularisation on $\mathbf{x}$, $\mathcal{P}$ takes the well-known form of a soft-thresholding operator.

\subsection{Deep unfolding}

In this work, instead of formalising regularizers analytically, we learn the proximal operator directly from an empirical data distribution, similar to \cite{mardani2018neural}. To that end, we unfold the iterations of \eqref{eq:iterations} into a K-layered neural network $U_\theta$, following \cite{gregor2010learning, monga2019algorithm, solomon2019deep} and shown in Fig. \ref{fig:network}. Unfolding allows us to learn both the proximal operator (under some parameterisation $\theta$) and the parameters of the gradient update, $\mathbf{V}^{(k)}$ and $\mathbf{W}^{(k)}$, in an end-to-end fashion.

Exploiting the block convolutional structure of $\mathbf{A}$, we implement $\mathbf{W}^{(k)}$ as:
\begin{equation}
    \mathbf{W}^{(k)}\mathbf{y} := \mathbf{w}_{1}^{(k)} \circledast \mathbf{y}_{1} + \mathbf{w}_{2}^{(k)} \circledast \mathbf{y}_{2} + ... + \mathbf{w}_{M}^{(k)} \circledast \mathbf{y}_{M},
    \label{eq:seven}
\end{equation}
and, likewise:
\begin{equation}
    \mathbf{V}^{(k)}\mathbf{\hat{x}}^{(k)} = \mathbf{v}^{(k)} \circledast \mathbf{\hat{x}}^{(k)},
    \label{eq:eight}
\end{equation}
where $\circledast$ denotes a convolutional operation, $\lbrace\mathbf{w}_{1}^{(k)}, ..., \mathbf{w}_{M}^{(k)}\rbrace$ and $\mathbf{v}^{(k)}$ are learned convolutional kernels, where we adopted a size of $3\times3$. 

We model $\mathcal{P}_{\theta}^{(k)}$ using a U-Net-style architecture with 9 convolutional layers with Leaky ReLU activations (see Table \ref{table:prox}, in appendix A for architectural details). As mentioned before, making $\mathcal{P}_{\theta}^{(k)}$ trainable avoids the need to devise hand-designed priors with an analytical proximal operator \cite{scherzer1993use}, and instead allows for data-driven proximal mappings that are learned from the training data distribution. As an additional advantage, unrolling the iterative scheme into a fixed-complexity trained computational graph yields a fast solution that avoids the problem of ambiguous computational complexity in iterative algorithms.

\section{Data Acquisition}\label{chapter:data_acquisition}



For this study we acquired both \textit{in-vivo} and \textit{in-silico} data. The \textit{in-vivo} data from 2 healthy volunteers was acquired using a Vantage system (Verasonics Inc., WA, USA), with the L11-4v and the L11-5v transducers. It was collected across 3 separate scan sessions, containing scans of the carotid artery in various alignments, muscle tissue, and the Achilles tendon. The \textit{in-silico} data consists of 20 randomly spaced point scatterers, at random distances from each other, created using the Verasonics simulation mode. Thirty of the \textit{in-silico} images also contained simulated speckle, made using 10,000 point scatterers of low reflectivity. 

An overview of the driving schemes and imaging parameters for both the LR input data and HR target data is given in Table~\ref{table:parameters}. The LR inputs (using 3 plane waves with a center frequency of 4 MHz) and HR targets (using 75 plane waves with a center frequency of 12 MHz) were acquired in successive frames. All 75 transmissions are equally spaced between the steering angles of $-18^{o}$ and $18^{o}$. All channel data was converted to beamformed RF using pixel-based beamforming. To further improve image quality in the HR acquisitions, we performed minimum variance (MV) beamforming, as opposed to regular delay-and-sum (DAS) for the LR acquisitions. To reduce motion between our input-target pairs, we gather the data using an acquisition sequence that gathers the low frequency image in the first frame, and the high frequency image in the subsequent frame.

\begin{table*}[t!]
\caption{Parameters of the US probes, driving schemes, and beamforming methods for the high-resolution (HR) targets and low-resolution (LR) inputs.}
\centering
    \begin{tabular}{c|c|c} 
        \textbf{Parameter} & \textbf{LR inputs} & \textbf{HR targets}  \\
        \hline
        Number of elements & 128 & 128 \\
        Array pitch & 0.3 mm & 0.3 mm \\
        Array aperture & 38.4 mm & 38.4 mm \\
        Transmit Frequency & 4 MHz & 12 MHz \\
        Number of plane waves & 3 ($-10.2^{o}$, $0^{o}$, and $10.2^{o}$) & 75 ($-18^{o}$ to $18^{o}$) \\
        Image reconstruction & pixel-based DAS & pixel-based MV + compounding
    \end{tabular}
\label{table:parameters}
\end{table*}


The training data contains 465 input-target pairs, one input being a stack of 3 LR single PW images, and the target being the corresponding HR image acquired subsequently. It contains a mix of \textit{in-vivo} and \textit{in-silico} data, at a ratio of approximately 3:1 (345 \textit{in-vivo} and 120 \textit{in-silico}). The validation set contains 60 input-target pairs, and the test set contains 45 input-target pairs (30 \textit{in-vivo} and 15 \textit{in-silico}).

In addition to this, due to the increased attenuation of the high-frequency targets, and therefore a lack of SNR at depth, we only include image pixels up to a set depth ($\sim$2.33 cm) for training. In addition, we exclude the first $\sim$0.3 cm, which predominantly consists of ringing. Note that for inference, we evaluate the full depth.  



\section{Training Strategy}\label{chapter:training}



\begin{figure*}[t!]
    \begin{subfigure}{\textwidth}
        \centering
        \includegraphics[width=\textwidth]{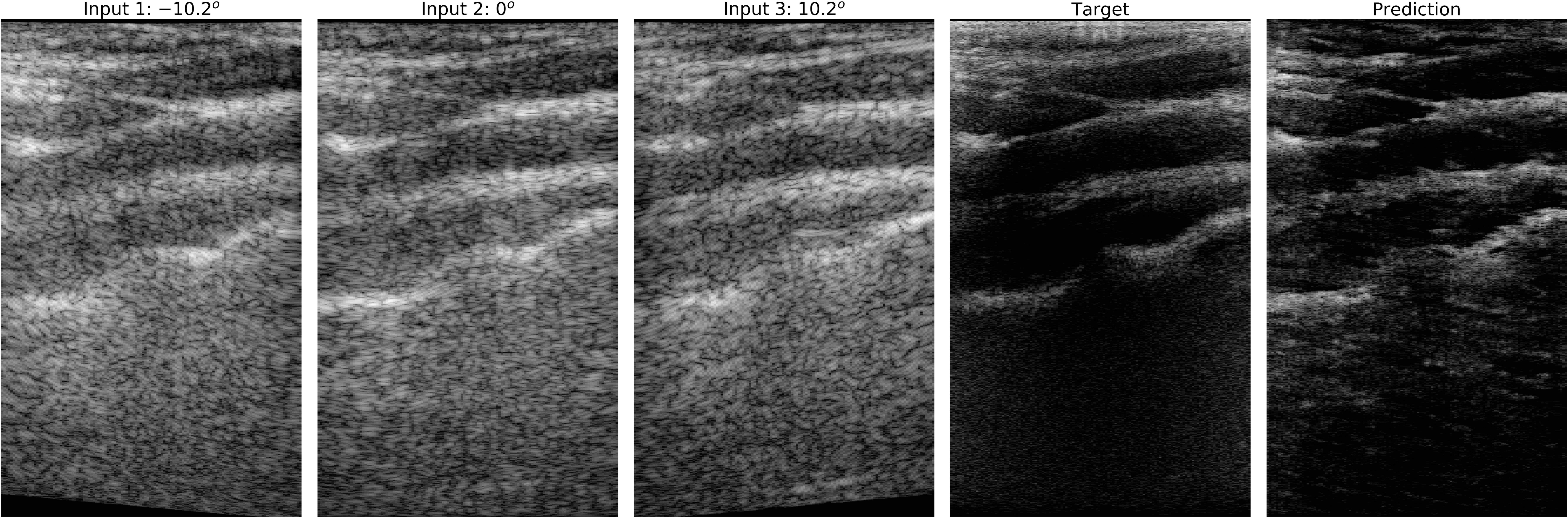}
        \label{fig:results_image_1}
    \end{subfigure}
    \vfill
    \begin{subfigure}{\textwidth}
        \centering
        \includegraphics[trim={0cm 0cm 0cm 8.8cm},clip,width=\textwidth]{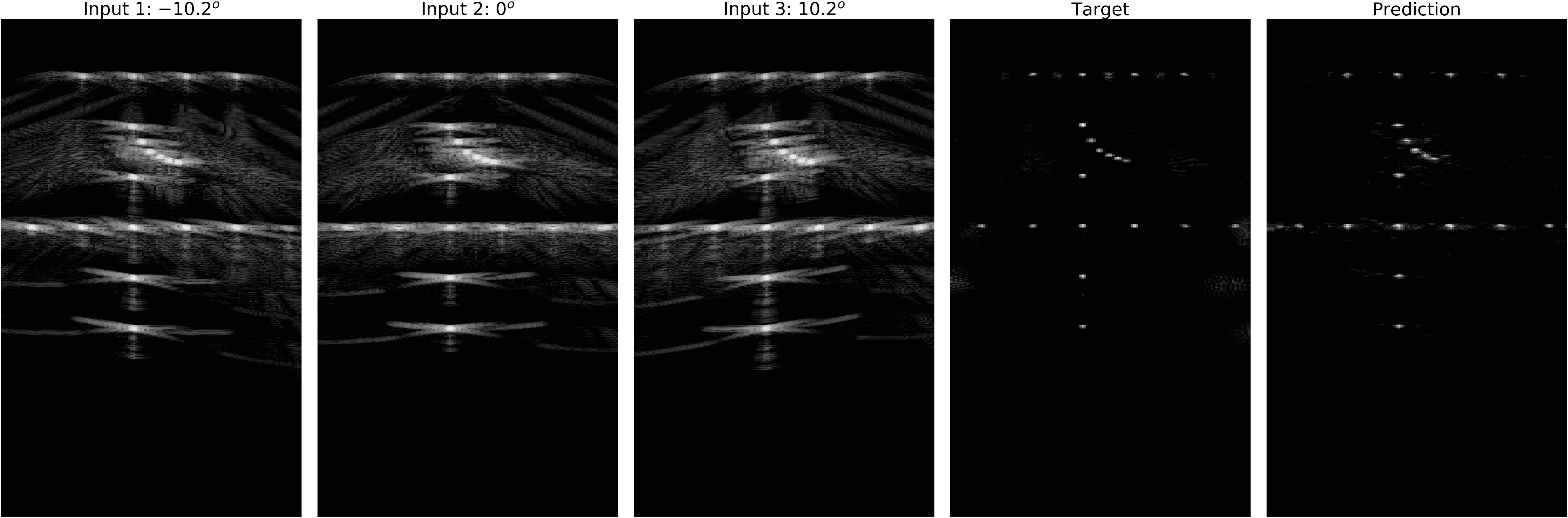}
        \label{fig:simulated_point}
    \end{subfigure}
    \caption{Log-compressed and envelope-detected set of 3 \textit{in-vivo} (first row) and \textit{in-silico} (second row) low-frequency input images, followed by the corresponding high frequency ground truth, compounded using 75 plane waves, and the deep network output. The first row depicts a longitudinal profile at the point of bifurcation of the carotid artery. All images were taken from the test set.}
    \label{fig:compare_all_res}
\end{figure*}

In this section, the training strategy is described in detail. To promote high fidelity in the image domain, we guide the optimization by adding an RF-to-image transformation function, $\mathcal{Q}$, to the computational graph before computing the losses. This way we emphasise a high resolution image in the image domain, after envelope detection and log-compression. We define $\mathcal{Q}$ as:
\begin{equation}
    \mathcal{Q}(\mathbf{x}) = 20\log_{10} \left(\epsilon + \frac{|Hilbert(\mathbf{x})|}{\max(|Hilbert(\mathbf{x})|)}\right),
    \label{eq:normer}
\end{equation} 
where $\epsilon$ is a small constant, and the values are clipped outside the range of interest (-60dB to 0dB). We combine 3 loss functions that capture various aspects of desired outputs: a pixel-wise distance loss, an adversarial loss, and a frame-to-frame loss (see Fig. \ref{fig:train} for an overview). 

\subsection{Pixel-Based Distance Loss}

We start with a pixel-based distance loss, which computes the average per-pixel $\ell_{1}$ distance between the reconstructed image $\mathcal{Q}(\hat{\mathbf{x}})= \mathcal{Q}(U_\theta(\mathbf{y}_t))$, and the target image $\mathcal{Q}(\mathbf{x})$: 
\begin{equation}
    \begin{aligned}
    \mathcal{L}_{\ell_{1}} &= \mathbb{E}_{(\mathbf{y}, \mathbf{x}) \sim p_{data}} \norm{\mathcal{Q}(\mathbf{x}_{t}) - \mathcal{Q}(U_\theta(\mathbf{y}_{t}))}_{1}
    \end{aligned}
\end{equation}
where $t$ indicates the frame number. Although we have framed the above in terms of expectation values, in practice, we approximate it by sampling a fixed dataset, i.e., the training data.

\subsection{Adversarial Loss}

We also include an adversarial loss to promote perceptually adequate outputs. Such adversaries are often used in GANs: deep generative models trained in a min-max game to map simple Gaussian-distributed noise vectors into outputs that follow a desired target distribution (e.g. an empirical distribution of natural images). Here we use a similar setup, and train our deep unfolded proximal gradient network in a min-max game with a neural adversary to produce outputs that, given the particular input data, fall within the distribution of high-resolution ultrasound images as sampled in our training set. By not just assessing outputs but rather input-output pairs, we mitigate hallucination of output image features that fall within the distribution of HR images, but are not consistent with the input data. 


Our neural adversary, $D_\psi$, outputs its belief that the image data spanned by its receptive field belongs to the set of high-resolution images. Here, $D_\psi$ is fully convolutional and has a receptive field smaller than the image dimensions (see Table \ref{table:discriminator}, in appendix A for architectural details), thus outputting probabilities for multiple distinct image regions (i.e. like a patchGAN \cite{isola2017image}). This directs the focus of the adversary towards local image quality features, rather than global semantic features. 


We train $D_\psi$ by minimizing the binary cross-entropy classification loss between its predictions and the labels (target or generated), evaluated on batches that contain both target images and generated images. At the same time, the deep compounding network $U_\theta$ is trained to maximize this loss, thereby attempting to fool the adversary. This leads to the following optimization problem across the training data distribution $p_{data}$:


            

\begin{multline}
    \begin{aligned}
        \hat{\theta}, \hat{\mathbb{\psi}} &= \argmax_{\mathbb{\psi}} \argmin_{\mathbb{\theta}} \left\{-\mathcal{L}_{adv}\right\} \\
        &= \argmax_{\mathbb{\psi}} \argmin_{\mathbb{\theta}} \mathbb{E}_{(\mathbf{y}_{t},\mathbf{x}_{t})\sim p_{data}} \\ &\qquad \left\{\log(D_{\psi}([\mathcal{Q}(\mathbf{y}_{t}),\mathcal{Q}(\mathbf{x}_{t})])) + \right. \\ & \left. \qquad \qquad \log(1-D_{\psi}([\mathcal{Q}(\mathbf{y}_{t}), \mathcal{Q}(U_\theta(\mathbf{y}_{t}))]))\right\},
    \end{aligned}
    \label{eqn:optimization_adversary}
\end{multline}
where $\theta$ are the parameters of the compounding network and $\psi$ are the parameters of the discriminator $D_\psi$. 






\begin{figure*}[t!]
    \begin{subfigure}{\textwidth}
        \centering
        \includegraphics[width=\textwidth]{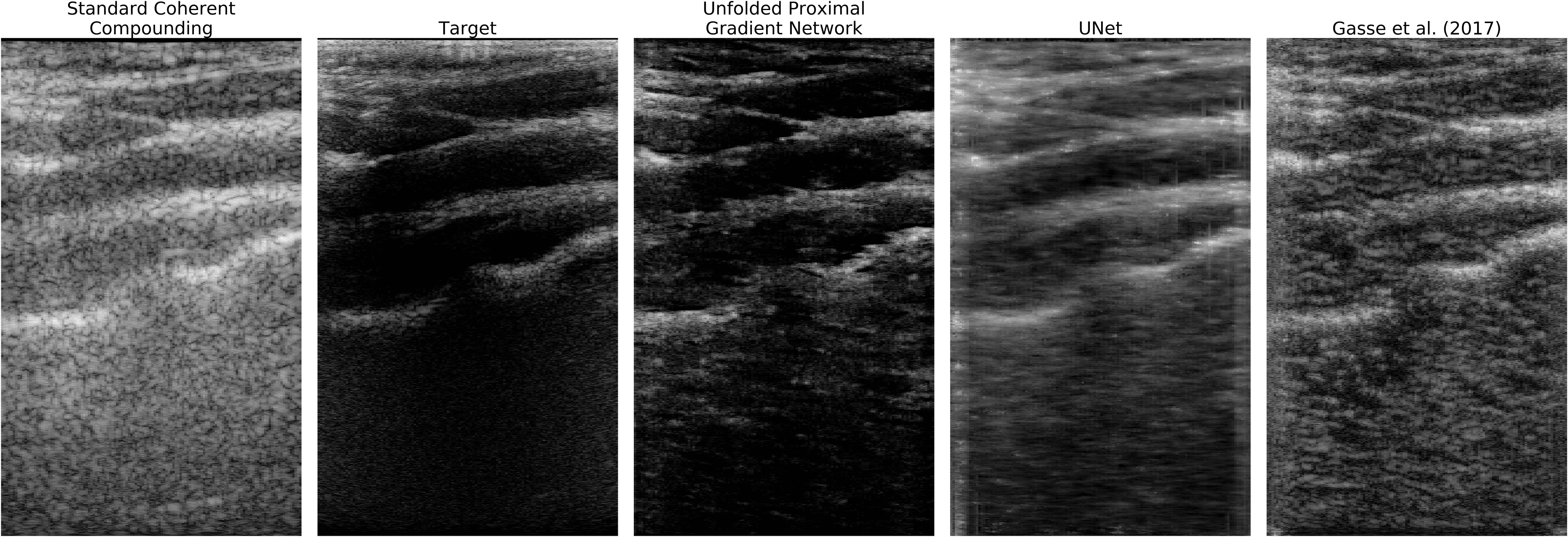}
        \label{fig:comparison_models}
    \end{subfigure}
    \vfill
    \begin{subfigure}{\textwidth}
        \centering
        \includegraphics[trim={0cm 0cm 0cm 8.8cm},clip,width=\textwidth]{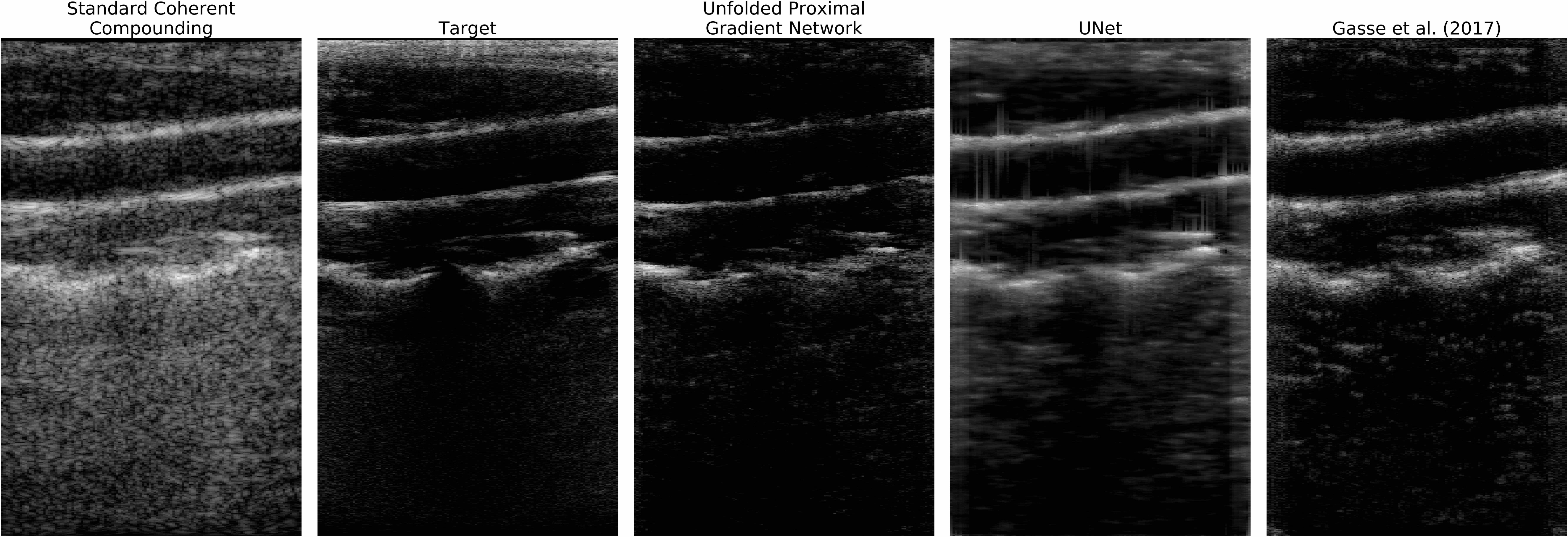}
        \label{fig:comparison_models_2}
    \end{subfigure}
    \caption{This figure compares different compounding methods against the target. We compare our approach against standard coherent compounding, compounding by a U-Net, and the architecture proposed by Gasse \textit{et al.} in \cite{gasse2017high}. The first row depicts a longitudinal profile of the carotid artery at the point of bifurcation. The second row depicts a longitudinal profile of the carotid artery. All images were taken from the test set.}
    \label{fig:compare_all_mods}
\end{figure*}

\subsection{Frame-to-Frame Loss}

In addition to the $\ell_{1}$ distance, and the adversarial loss mentioned above, we also implement an $\ell_{2}$ distance on consecutive frames ($(t-1)^{th}$ and $t^{th}$ frame) as predicted by $U_\theta$. This is done to promote consistency between frames while suppressing (frame-wise) uncorrelated noise:
\begin{equation}
    \mathcal{L}_{ftf} = \mathbb{E}_{(\mathbf{y}_{t}, \mathbf{y}_{t-1}) \sim p_{data}} \norm{\mathcal{Q}(U_\theta(\mathbf{y}_{t-1})) - \mathcal{Q}(U_\theta(\mathbf{y}_{t}))}_{2}^{2} .
\end{equation}

\subsection{Training details and parameters}

The total loss function is a weighted sum of the aforementioned losses, and given by:
\begin{equation}
    \mathcal{L}_{tot} = \lambda_{1} \cdot \mathcal{L}_{adv} + \lambda_{2} \cdot \mathcal{L}_{\ell_{1}} + \lambda_{3} \cdot \mathcal{L}_{ftf}
    \label{eq:final_simplified}
\end{equation}
where $\lambda_{i}$ are weight terms, with $\lambda_{1} = 1.0$, $\lambda_{2} = 1.0$, and $\lambda_{3} = 0.001$ in our experiments, determined empirically.

We train the network in a greedy manner \cite{bengio2007greedy, belilovsky2019greedy}, i.e. fold-by-fold. In the first iteration, we only train the first fold. We then add the next fold, freezing the first, and continue training. This process continues until all folds are trained. Note that the last convolutional layer of the architecture (see Fig. \ref{fig:train}) is trained in every iteration. Each fold is trained for 1800 epochs, with a batch size of 1. We use the Adam optimiser \cite{kingma2014adam} with a learning rate of $1 \times 10^{-5}$, and with a $\beta_{1} = 0.5$. All other optimiser parameters are left at default, as described in \cite{kingma2014adam}.

All of the networks were created using Python 3 and TensorFlow 2 \cite{tensorflow2015-whitepaper}, and trained using an NVIDIA 2080 Ti GPU.

\section{Results}\label{chapter:results}

\begin{figure}
    \centering
    \includegraphics[width=0.45\textwidth]{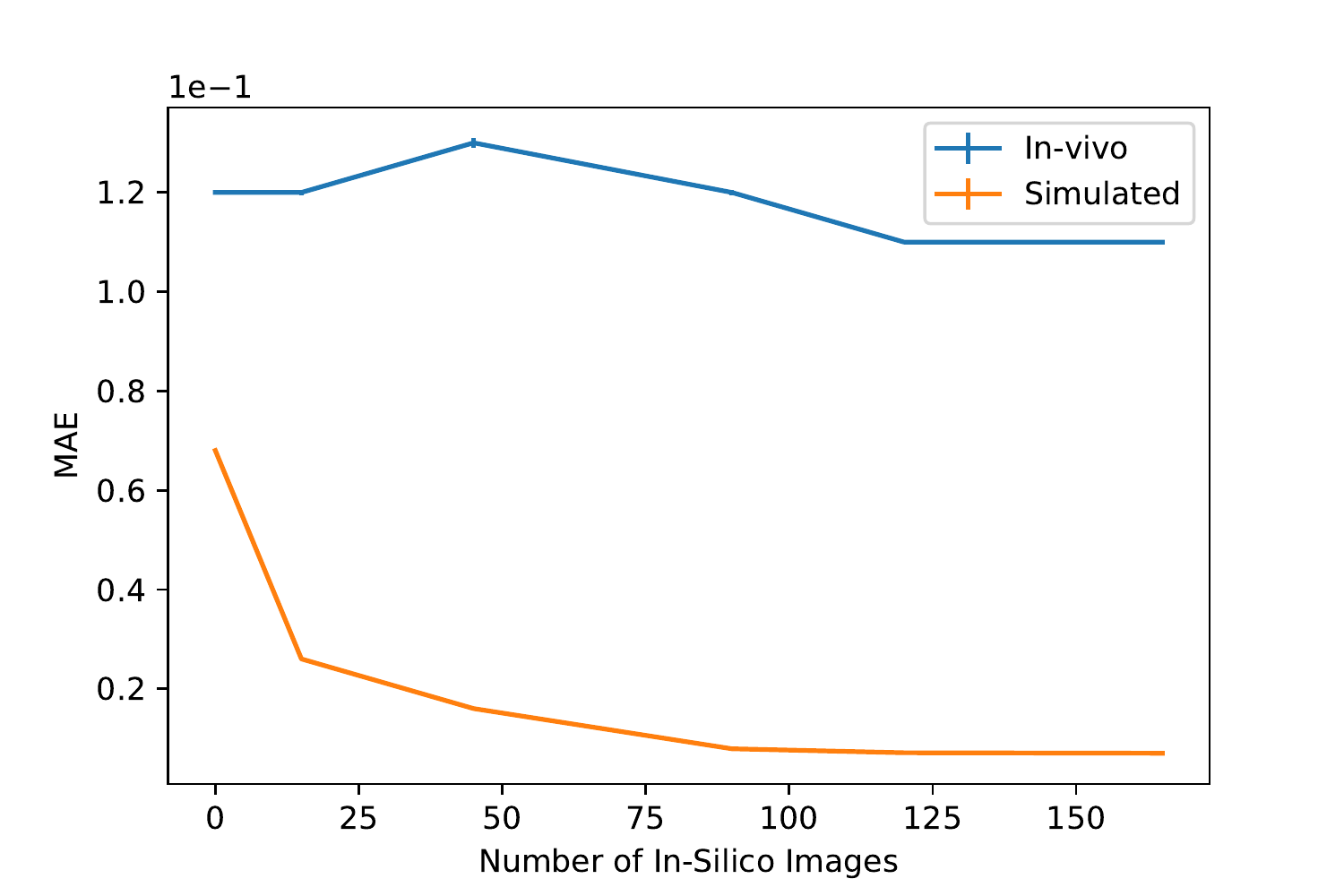}
    \caption{This figure plots the performance of the proposed network as the volume of \textit{in-silico} data in the training set in increased. The performance on the \textit{in-vivo} and \textit{in-silico} test sets, are shown separately. Notably, \textit{in-vivo} performance improves when adding a significant amount of \textit{in-silico} images to the train set. Due to vast differences in the distributions of \textit{in-silico} and \textit{in-vivo} data, we plot the trends separately such that trend lines are comprehensible for both \textit{in-vivo} and combined test sets.}
    \label{fig:simulated_data}
\end{figure}

\begin{figure}[t!]
    \centering
    \includegraphics[width=0.4\textwidth]{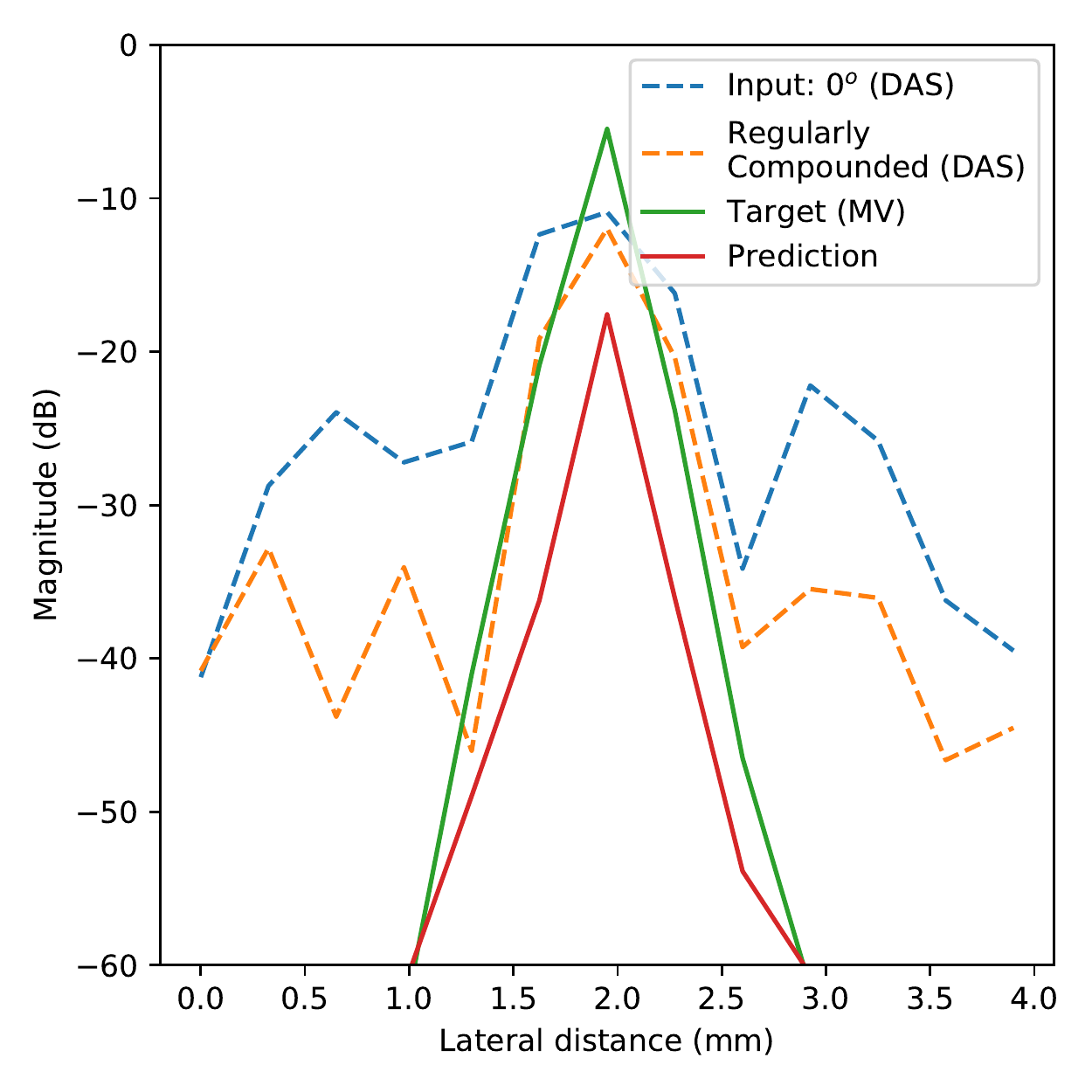}
    \caption{The graphs displays the beam profiles of a point scatterer generated by a \textit{in-silico}, and the subsequent prediction on that data.}
    \label{fig:fwhm_im}
\end{figure}

\begin{figure}
    \centering
    \includegraphics[width=0.45\textwidth]{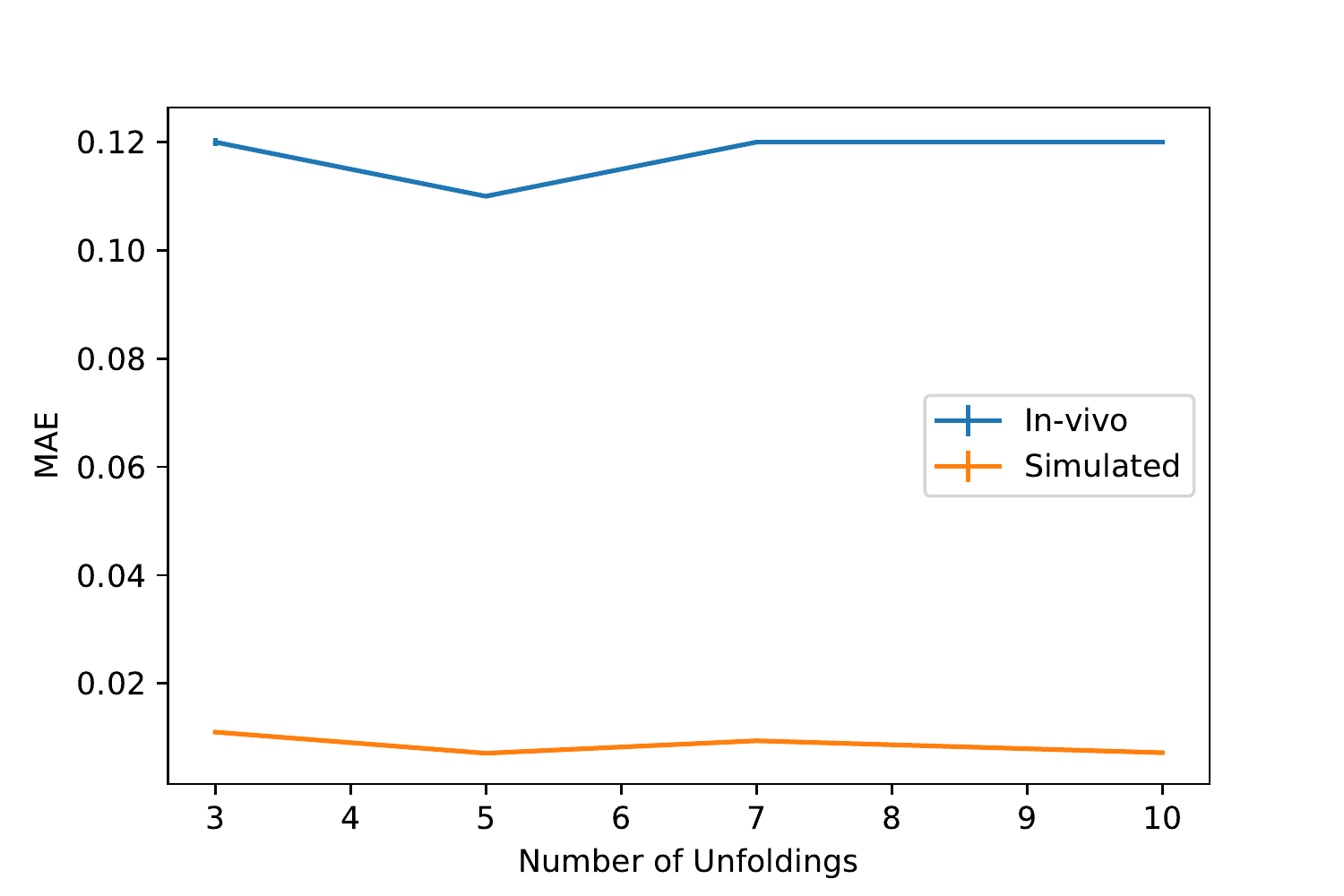}
    \caption{The graph displays the relationship between the number of unfoldings in the deep proximal network, and the performance on the test set. Due to vast differences in the distributions of \textit{in-silico} and \textit{in-vivo} data, we plot the trends separately such that trend lines are comprehensible for both \textit{in-vivo} and combined test sets.}
    \label{fig:unfolded_res}
\end{figure}

The results presented in this section were obtained using a network with 5 folds (unless otherwise stated) using the aforementioned framework to construct and train an unfolded proximal gradient network.


The first row of Fig. \ref{fig:compare_all_res} shows the log-compressed and envelope detected input images, obtained using single 4 MHz transmits, along with the corresponding image target acquired using 75 PW transmits at 12 MHz, followed by the compounded image obtained using the proposed method. Perceptually, the resulting image has a similar fidelity as the target, showing high structural resolution and contrast.  

In the second row of Fig. \ref{fig:compare_all_res}, we display the result of our unfolded proximal gradient network when applied to \textit{in-silico} point scatterers, showing strong contrast, resolution, and side-lobe suppression. Fig. \ref{fig:fwhm_im} displays a comparison of the beam profile of a point scatterer, between different methods of beamforming and compounding methods.

Figure \ref{fig:compare_all_mods} show a comparison between the various techniques of compounding the given set of LR images. We compare the resulting images from standard coherent compounding, the target (MV), the prediction by the proposed network, a UNet, and the CNN by Gasse et al. \cite{gasse2017high}. 

Table \ref{table:results} quantifies the Mean Absolute Error (MAE) values, the number of parameters, and the un-optimised inference times of the proposed network (with 5 folds), a UNet-like network (UNet Lite; see Table \ref{table:unet_lite} in appendix A for architectural details) with a similar number of parameters, a full UNet \cite{ronneberger2015u}, and the CNN of \cite{gasse2017high}, containing about 60\% more free parameters than the proposed network. We chose UNets as the benchmark because it is shown to be a very good baseline for such problems \cite{hauptmann2020unreasonable}. Furthermore, most other works rely on a variation of a neural network with skip connections, so the UNet is a good common denominator. This comparison is made on the test data that was gathered for use in this project, as described in Section \ref{chapter:data_acquisition}. The benchmark networks were trained using the same training strategy as the proposed network, albeit not in a greedy training strategy. 

Table \ref{table:ablation} shows the results of an ablation study evaluating the contribution of the individual loss terms in \eqref{eq:final_simplified} to the results. In both Table \ref{table:results} and Table \ref{table:ablation}, we separate the calculation of the MAE between the \textit{in-vivo} and \textit{in-silico} test samples. 


Additionally, we performed experiments to ascertain the volume of \textit{in-silico} data that should be added to the training loop in order to obtain the best performance. Fig. \ref{fig:simulated_data} quantifies the result of such experiments, showing the MAE metric as a function of the number of \textit{in-silico} training data to \textit{in-silico} training data. We also performed an empirical search for the optimal number of layers, and trained the network with 3, 5, 7, and 10 folds. The result of this search is shown in Fig. \ref{fig:unfolded_res}, which displays the MAE metric as a function of the number of unfoldings of the proximal network. 

        
        
        

\begin{table*}[htbp]
\caption{The table displays a comparison made between the proposed Unfolded Proximal Gradient Scheme, U-Net Lite (a network with a similar number of parameters), a full UNet, and the CNN published by Gasse et al. (2017). We compare the Mean Absolute Error (MAE) (including mean and variance across images in the test set), the number of parameters, and inference times (tested on an NVIDIA 3080 Ti).}
\centering
    \begin{tabular}{c|c|c|c|c} 
        \multirow{2}*{\textbf{Network}} & \multicolumn{2}{|c|}{\textbf{Mean Absolute Error (mean $\pm$ var)}} & \textbf{No. of} & \textbf{Inference} \\\cline{2-3}
         & In-Vivo & In-Silico & \textbf{Parameters} & \textbf{Time} \\ 
        \hline
        Unfolded Proximal & \multirow{2}*{1.1$\times 10^{-1}$ $\pm$ 2.6$\times 10^{-4}$} & \multirow{2}*{7.1$\times 10^{-3}$ $\pm$ 1.48$\times 10^{-6}$} & \multirow{2}*{$\sim$ 175 k} & \multirow{2}*{$\sim$ 20 ms} \\
        Gradient &  &  &  & \\
        
        UNet & 1.16$\times 10^{-1}$ $\pm$ 9.8$\times 10^{-8}$ & 1.1$\times 10^{-2}$ $\pm$ 9.7$\times 10^{-7}$ & $\sim$ 1175 k & $\sim$ 8 ms \\
        
        UNet Lite & 1.34$\times 10^{-1}$ $\pm$ 1.93$\times 10^{-3}$ & 1.5$\times 10^{-2}$ $\pm$ 4.94$\times 10^{-6}$ & $\sim$ 160 k & $\sim$ 6 ms \\
        
        CNN by & \multirow{2}*{1.4$\times 10^{-1}$ $\pm$ 1.96$\times 10^{-3}$} & \multirow{2}*{1.8$\times 10^{-2}$ $\pm$ 5.65$\times 10^{-6}$} & \multirow{2}*{$\sim$ 280 k} & \multirow{2}*{$\sim$ 6 ms}\\
        Gasse et al. (2017) &  &  &  & \\
    \end{tabular}
\label{table:results}
\end{table*}

\begin{table*}[htbp]
\caption{The following table shows an ablation study, evaluating the contribution of loss terms as given in equation \eqref{eq:final_simplified}. We present the Mean Absolute Error (MAE), reporting both its mean and standard deviation across the test set.}
\centering
    \begin{tabular}{c|c|c|c|c|c} 
        \multicolumn{3}{c|}{\textbf{Losses}} & \textbf{Training} & \multicolumn{2}{c}{\textbf{Mean Absolute Error (mean $\pm$ var)}} \\ 
        \hline
        Pixel-based distance & Adversary & Frame-to-Frame & & In-Vivo & In-Silico \\ 
        \hline
        \checkmark & \checkmark & \checkmark & Greedy & 1.1$\times 10^{-1}$ $\pm$ 2.6$\times 10^{-4}$ & 7.1$\times 10^{-3}$ $\pm$ 1.48$\times 10^{-6}$
        \\
        - & \checkmark & \checkmark & Greedy & 1.3$\times 10^{-1}$ $\pm$ 1.03$\times 10^{-3}$ & 7.4$\times 10^{-3}$ $\pm$ 1.18$\times 10^{-6}$
        \\
        \checkmark & - & \checkmark & Greedy & 1.03$\times 10^{-1}$ $\pm$ 4.79$\times 10^{-4}$ &7.34$\times 10^{-3}$ $\pm$ 1.16$\times 10^{-6}$
        \\        
        \checkmark & \checkmark & - & Greedy &1.12$\times 10^{-1}$ $\pm$ 2.48$\times 10^{-4}$ & 7.43$\times 10^{-3}$ $\pm$ 1.19$\times 10^{-6}$

    \end{tabular}
\label{table:ablation}
\end{table*}





We also re-trained the network after the initial greedy training process, but received no appreciable increase in the quantitative or qualitative metrics of the consequent results.

\section{Discussion}\label{chapter:discussion}




We first start with Fig. \ref{fig:simulated_data}, which makes the distinction between testing the network with \textit{in-silico}, and \textit{in-vivo} samples. Due to the stark differences in the distribution of the images, the variance of the combined test set can be quite high. Therefore, we plot two trend lines, showing that results improve for both \textit{in-vivo} images, and simulated point scatterer data. This indicates that adding \textit{in-silico} data in the training loop can also help improve the quality of not just \textit{in-silico} inputs, but also \textit{in-vivo} data. It also shows that augmenting limited \textit{in-vivo} training data with \textit{in-silico} data is a valid approach. 

Fig. \ref{fig:fwhm_im} shows that the proposed network does indeed improve the beam profile of point scatterers. The network not only manages to improve the lateral localisation of the point scatterer, but also reduces the noise in its periphery, as seen in the input profile and the regularly compounded profile.

While augmenting the \textit{in-vivo} dataset with a sizeable volume of \textit{in-silico} samples improves results, eventually hitting a point of diminishing returns, the same cannot be said of adding folds to the network. In Fig. \ref{fig:unfolded_res}, we show that increasing the number of folds in a network does not necessarily lead to better results. But rather that the optimum number of folds may lay with an intermediary number of unfoldings, which is in line with other works that deal with unfolded networks \cite{solomon2019deep}.

In Fig. \ref{fig:compare_all_res}, we see that the network does succeed in improving the spatial resolution and contrast of the input signal, with the image quality approaching the ground truth. We also have predictions by the benchmark networks compared to the unfolded network shown in Fig. \ref{fig:compare_all_mods}. Although the UNet performs quite well when considering the metrics from Table \ref{table:results}, visually, we see that the results are hazy and undefined. Whereas the CNN is noisy, and is potentially hallucinating the noise that we observe in the predicted image. We can therefore hypothesise that the UNet based networks are perhaps under-specified, and have therefore learnt a shortcut to optimise the problem at hand. While the CNN performs better than the UNet visually, it comes at the cost of a larger computational cost, and a more noticeable noise profile.

A similar effect is in action in Table \ref{table:ablation}, when $\lambda_{1} = 0.0$ (adversarial loss weight term). The MAE for the \textit{in-vivo} samples indicate that the results are better, however, qualitatively/perceptual performance is worse. This is because the adversarial loss tends to promote reconstruction of the high frequency, crisp details of an image, while the $\ell_{1}$ loss is in practice dominated by errors in the low frequency components. Thus, without a perceptual penalty on the sharpness of an image (via the adversary), the resulting image tends to be blurry and hazy. Fig. \ref{fig:comparison_models_3}, in appendix B displays the images that are obtained from networks that have been trained with the $\mathcal{L}_{\ell_{1}}$ and the $\mathcal{L}_{adv}$ components turned off individually. 



While we employ a combination of both pixel-based and distribution-based loss, the neural network optimises for the lowest value of loss, rather than the best visual representation of the US signal data. Although it is advantageous to use such a combination, it presents us with a problem that has been encountered in US imaging before, and demonstrates why it is equally important to visually inspect the resulting images in addition to relying on metrics. More importantly, the benchmark networks lack any of the physics-based design considerations that the proposed network possesses.


These results allude to the advantages of using model based networks in challenging ill-posed problems, as it appears to promote stability by virtue of its architecture. We also find that it is crucial to transform the output of the unfolded network using an RF-to-Image transformation function (equation \eqref{eq:normer}) before the calculation of the losses (much more trivial to balance loss terms), or passing it on to the discriminator.




\section{Conclusions}\label{chapter:conclusion}

We proposed a model-based deep neural architecture for high-resolution plane wave compounding, incorporating strong physics-based inductive biases. Following the unfolding approach, we observe that the proposed network out-performs several benchmarks by approximately 5\%, using the MAE metric.  

Perceptually, the proximal network produces images with enhanced fidelity, containing high structural resolution and contrast, as shown in Figs. \ref{fig:compare_all_res} and \ref{fig:compare_all_mods}. We also derive a high quality prediction on the point scatterers as seen in the second row of Fig. \ref{fig:compare_all_res}, with a high degree of side-lobe suppression, resolution, and contrast. The same is confirmed using a 1D lateral profile of the point scatterers. We also further demonstrate that it is beneficial to include a significant number of \textit{in-silico} samples in the training of the network, and that increasing the number of folds beyond 5 does not necessarily lead to better performance, which is in line with other works that have trained unfolded networks \cite{solomon2019deep}. Replacing the proximal operator with other potential candidates is a great avenue for further research, along with enhanced data augmentation.  

While it is typically infeasible to make use of MV beamforming in real-time applications, we have presented a neural architecture that can greatly enhance the resolution and contrast of ultrasound images, over DAS beamformed images, with great potential for real-time applications. We achieve the above results with a relatively low parameter count when compared to other architectures. We conclude that the model-based approach proposed here can produce better results under a comparatively small computational foot-print. 








\section*{Acknowledgment}

The authors would like to thank Dr. Elik Aharoni, Ronnie Rosen, and Oded Drori for their guidance and help in collecting the \textit{in-vivo} data. We would also like to thank the Weizmann Institute of Science for accommodating a visit in the spirit of international collaboration.  

\bibliographystyle{IEEEtran}
\bibliography{refs}

\section*{Appendix A}\label{chapter:appendix}

\noindent Given below are the details of the neural network architectures used in this work. Table \ref{table:prox} details on the architecture of the adopted proximal network, Table \ref{table:discriminator} gives the architecture of the discriminator, and Table \ref{table:unet_lite} provides the UNet Lite network used in the benchmarks. 

\begin{table}[h!]
\caption{The architectural design of the neural proximal operator.}
\begin{tabular}{||c c c c c||}
\hline
Layer \# & Type          & Kernel Size & Activation & Output Shape \\ \hline \hline
1 & Convolutional & 3x3 & LeakyRelu & (2048x128x1) \\ \hline
2 & Convolutional & 3x3 & LeakyRelu & (2048x128x8) \\ \hline
3 & Convolutional & 3x3 & LeakyRelu & (2048x128x16)\\ \hline
4 & Convolutional & 3x3 & LeakyRelu & (1024x64x32) \\ \hline
5 & Convolutional & 3x3 & LeakyRelu & (1024x64x32) \\ \hline
6 & Dropout(0.3)  & -   & -         & (1024x64x32) \\ \hline
7 & Concatenate   & -   & -         & (2048x128x40) \\ \hline
8 & Convolutional & 3x3 & LeakyRelu & (2048x128x16) \\ \hline
9 & Convolutional & 3x3 & LeakyRelu & (2048x128x32) \\ \hline
10& Convolutional & 3x3 & LeakyRelu & (2048x128x1) \\ \hline
\end{tabular}
\label{table:prox}
\end{table}

\begin{table}[h!]
\caption{The architectural design of the neural discriminator used for adversarial training.}
\begin{tabular}{||c c c c c||}
\hline
Layer \# & Type          & Kernel Size & Activation & Output Shape \\ \hline \hline
1 & Concatenate   & -   & -         & (2048x128x4) \\ \hline
2 & Convolutional & 3x3 & LeakyRelu & (1024x64x32) \\ \hline
3 & Convolutional & 3x3 & LeakyRelu & (512x32x64)  \\ \hline
4 & Convolutional & 3x3 & LeakyRelu & (256x16x128) \\ \hline
5 & Convolutional & 3x3 & Sigmoid   & (256x16x1)   \\ \hline
\end{tabular}
\label{table:discriminator}
\end{table}

\begin{table}[h!]
\caption{The architectural design of the baseline Unet Lite.}
\begin{tabular}{||c c c c c||}
\hline
Layer \# & Type          & Kernel Size & Activation & Output Shape \\ \hline \hline
1  & Concatenate   & -   & -         & (2048x128x3) \\ \hline
2  & Convolutional & 3x3 & LeakyRelu & (2048x128x16)\\ \hline
3  & Convolutional & 3x3 & LeakyRelu & (2048x128x32)\\ \hline
4  & Convolutional & 3x3 & LeakyRelu & (1024x64x64)\\ \hline
5  & Convolutional & 3x3 & LeakyRelu & (1024x64x64)\\ \hline
6  & Convolutional & 3x3 & LeakyRelu & (2048x128x64)\\ \hline
7  & Dropout(0.3)  & -   & -         & (2048x128x64) \\ \hline
8  & Concatenate   & -   & -         & (2048x128x96)\\ \hline
9  & Convolutional & 3x3 & LeakyRelu & (2048x128x60)\\ \hline
10 & Convolutional & 3x3 & LeakyRelu & (2048x128x16)\\ \hline
11 & Convolutional & 3x3 & LeakyRelu & (2048x128x1)\\ \hline
\end{tabular}
\label{table:unet_lite}
\end{table}

\section*{Appendix B}\label{chapter:appendixB}

\begin{figure*}[t!]
    \centering
    \includegraphics[width=\textwidth]{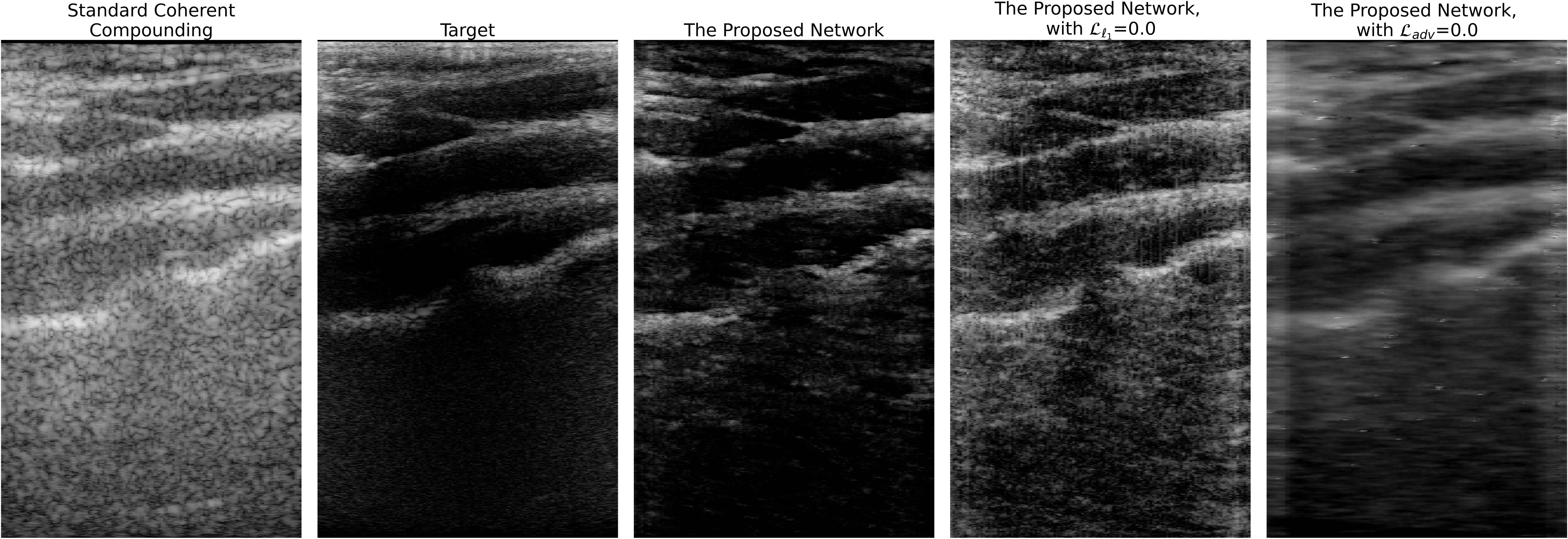}
    \caption{The images show the effect of switching off parts of the total objective function on the final prediction. It displays standard coherent compounding, the target image, the proposed network, and two other iterations of the proposed network where $\mathcal{L}_{\ell_{1}}$ and $\mathcal{L}_{adv}$ are set to zero, individually. The anatomy is a longitudinal profile of the carotid artery. These images have been taken from the test set.}
    \label{fig:comparison_models_3}
\end{figure*}

\noindent Figure \ref{fig:comparison_models_3} qualitatively displays the effect of the individual components of the total loss function. It shows us that a lack of the $\mathcal{L}_{adv}$ term tends to provide results that do not have highly defined features. On the other hand, a prediction made using a network trained without the $\mathcal{L}_{\ell_{1}}$ term yields results that lack contrast.

\end{document}